\documentclass[conference]{IEEEtran}
\IEEEoverridecommandlockouts

\usepackage{cite}
\usepackage{amsmath,amssymb,amsfonts}
\usepackage{graphicx}
\usepackage{textcomp}

\ifCLASSOPTIONcompsoc
  \usepackage[caption=false,font=normalsize,labelfont=sf,textfont=sf]{subfig}
\else
  \usepackage[caption=false,font=scriptsize]{subfig}
\fi

\usepackage{url}
\usepackage{graphicx}
\usepackage[table,xcdraw]{xcolor}
\usepackage{amsthm}
\usepackage{svg}
\usepackage{bm}
\usepackage{algorithm}
\usepackage[noend]{algpseudocode}
\usepackage{setspace}
\usepackage{mwe}

\usepackage[OT1,T1]{fontenc}
\usepackage[scaled=.88]{berasans}

\DeclareMathOperator*{\argmin}{argmin}

\algdef{SE}[DOWHILE]{Do}{doWhile}{\algorithmicdo}[1]{\algorithmicwhile\ #1}%

\algrenewcommand\algorithmicindent{1em}%

\def\BibTeX{{\rm B\kern-.05em{\sc i\kern-.025em b}\kern-.08em
    T\kern-.1667em\lower.7ex\hbox{E}\kern-.125emX}}

\begin{document}

\title{Model-Based Reinforcement Learning Framework of Online Network Resource Allocation}

\author{
\IEEEauthorblockN{
	Bahador Bakhshi,
	Josep Mangues-Bafalluy
}
\IEEEauthorblockA{Centre Tecnologic de Telecomunicacions de Catalunya (CTTC/CERCA), Spain\\
\{bbakhshi, josep.mangues\}@cttc.cat}

\thanks{{\color{black}This  work  has  been  partially  funded  by  the EC  H2020  5Growth  Project  (grant  no.  856709),  and Generalitat de Catalunya grant 2017 SGR 1195.}}
}

\maketitle

\begin{abstract}
Online Network Resource Allocation (ONRA) for service provisioning is a fundamental problem in communication networks. As a sequential decision-making under uncertainty problem, it is promising to approach ONRA via Reinforcement Learning (RL). But, RL solutions suffer from the \emph{sample complexity} issue; i.e., a large number of interactions with the environment needed to find an efficient policy. This is a barrier to utilize RL for ONRA as on one hand, it is not practical to train the RL agent offline due to lack of information about future requests, and on the other hand, online training in the real network leads to significant performance loss because of the sub-optimal policy during the prolonged learning time. This performance degradation is even higher in non-stationary ONRA where the agent should continually adapt the policy with the changes in service requests. To deal with this issue, we develop a general resource allocation framework, named RADAR, using model-based RL for a class of ONRA problems with the known immediate reward of each action. RADAR improves sample efficiency via exploring the state space in the background and exploiting the policy in the decision-time using synthetic samples by the \emph{model} of the environment, which is trained by real interactions. Applying RADAR on the multi-domain service federation problem, to maximize profit via selecting proper domains for service requests deployment, shows its continual learning capability and up to 44\% performance improvement w.r.t. the standard model-free RL solution.
\end{abstract}

\begin{IEEEkeywords}
Online Resource Allocation, Model-based Reinforcement Learning, Service Federation, Continual Learning
\end{IEEEkeywords}

\vspace{-3mm}
\section{Introduction}
\label{sec:introduction}
Resource allocation is a fundamental problem in communication networks \cite{xu2021survey, yang2020recent}, wherein network resources are allocated for service requests to optimize an objective function, e.g., the service provider's profit, subject to satisfying Quality of Service (QoS) requirements. Resource allocation problems are categorized into two categories, namely online and offline. In the former, service requests arrive one-by-one over time, and the Resource Allocator (RA) allocates resources for each request without knowledge of future ones. But in offline problems, it is assumed that all requests are given/known in advance. In many practical use cases, where the exact information of future requests is not available, online RA solutions are the de facto approach.

Due to the central role of the resource allocation problem, various solutions have been developed over the last decades including problem-specific heuristic RA algorithms, e.g., QoS routing \cite{guck2017unicast}, game theory and meta-heuristic based solutions \cite{rahmani2020game}, and various optimization theory techniques \cite{ejaz2020comprehensive}. Recently, by the emergence of Machine Learning (ML) and its successful applications, ML-based solutions for network resource allocation and management have also been proposed \cite{morocho2019machine}. Reinforcement Learning (RL) is a class of ML solutions that contrary to supervised learning methods, learns the suitable action in each state, known as policy, via the rewards it receives from interactions with the environment \cite{sutton2018reinforcement}. RL is an efficient tool to approach a class of problems known as \emph{sequential decision making under uncertainty} (SDMU), where the decision-maker needs to take actions without knowing the uncertain future. In applying RL to SDMU, the decision-maker is an RL agent that in a trial-and-error process explores the state space, and eventually finds an appropriate policy. 

Online Network Resource Allocation (ONRA) problems are indeed instances of SDMU as the RA should allocate resources while the future requests are uncertain. So, RL is a promising approach for ONRA problems \cite{bakhshi2021globe, xiong2020resource, ye2019deep}. Although the main advantage of RL is that it does not need explicitly labeled training data, to find an efficient policy, the agent requires a huge number of interactions with the environment, which is known as the \emph{sample complexity} issue \cite{moerland2020model}; for example, in the well-known deep RL solution for playing Atari games \cite{mnih2015human}, the agent is trained using $5\times10^{7}$ frames of the game, or in our proposed solution for multi-domain service federation (MDFS) \cite{bakhshi2021globe}, it takes about 9 $\times 10^5$ interactions to find the near-optimal policy.
In applying RL to ONRA, where the agent learns the policy in an online manner in the real network, sample complexity can cause significant performance degradation because before finding the (near) optimal policy, agent's decisions are sub-optimal for a high number of real requests. Dealing with the RL sample complexity in ONRA  is the research gap that we aim to address in this paper. 

\emph{Model-Based} RL (MBRL) is an approach to improve sample efficiency of RL by integrating a model of the environment in the RL \cite{moerland2020model}. The core idea is that in each interaction with the real environment, besides improving the policy, the agent also learns a model of the environment. Then it generates synthetic interactions by this model which are used to improve the policy. MBRL has been used in contexts like playing Atari games \cite{kaiser2019model}, and robotic \cite{polydoros2017survey}; recently, it has also got attention in communication networks. In \cite{yang2019miras}, an MBRL approach was developed for microservice-based applications resource allocation. Dynamic computational resource allocation in the cloud was studied in \cite{chen2020reinforcement}, where the problem is formulated as a Markov Decision Process (MDP) and by the value iteration algorithm, an MBRL method was developed. In \cite{jiang2021resource}, again resource allocation in the cloud was investigated, the authors developed an MBRL approach to satisfy the application constraints on the rate of allocation changes. These solutions are not applicable to ONRA as they are specifically designed for cloud environment to model computational resources and workloads. In this paper, we develop a \emph{general framework} for ONRA using MBRL, and make the following contributions:
\begin{itemize}
    \item a general framework, called RADAR, with the capability of continual learning is developed for ONRA problems;
    \item by exploiting the flexibility of the framework, four algorithms are implemented to be used in different use cases;
    \item as a proof of concept, the RADAR-based algorithms are applied and evaluated in the MDSF problem.
\end{itemize}

The remainder of this paper is organized as follows. In Section \ref{sec:system_problem}, the system model and the sample complexity issue are elaborated in more detail. The RADAR framework is presented in Section \ref{sec:framework}. The MDSF problem is solved by RADAR in Section \ref{sec:md_federation}, which is numerically evaluated in Section \ref{sec:results}. Finally, Section \ref{sec:conclusion} concludes this paper.

\vspace{-0.7mm}
\section{System Model and Problem Statement}
\label{sec:system_problem}
\begin{figure}[t]
\begin{center}
\centerline{\includegraphics[width=0.95\linewidth]{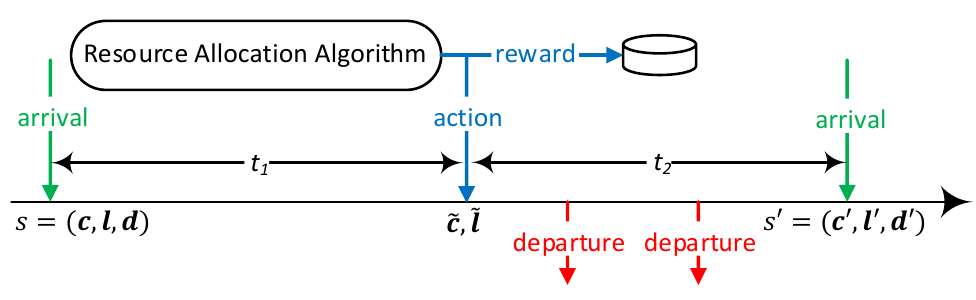}}
\vspace{-0.4cm}
\caption{The timeline of the KRORA problems}
\label{fig:krora}
\vspace{-6.5mm}
\end{center}
\end{figure}

\vspace{-1.25mm}
In this paper, we study a class of ONRA problems, named Known-Reward Online Resource Allocation (KRORA); which is depicted in Fig.~\ref{fig:krora}, and identified by the following characteristics. First, the problem is online; i.e., requests for the services defined in the service catalog arrive one-by-one. For each request, without knowledge of future requests, the RA algorithm runs, and after time $t_{1}$, takes a resource allocation action. Second, each  RA action has a \emph{known} immediate reward, as in practice  the service provider either knows or can determine the immediate consequences of each RA action. This is exemplified in the following paragraph.  Third, RA's objective is to optimize a cumulative long-term reward. Fourth, the next request arrives after time $t_{2}$. Meanwhile, during this period, some alive services terminate and depart the network. If an arrival or departure occurs in the time interval $t_{1}$, it will be taken into account as soon as the RA algorithm ends. 

A large number of practical ONRA problems are indeed in the KRORA class as they have the mentioned characteristics.
For example, in the online flow routing problem, each request is a demand for a flow from a source node to a destination node with known predefined QoS requirements. The objective is to maximize the total number of accepted flows; so, the immediate reward is 1.
Another example is the MDSF problem, which is elaborated in Section \ref{sec:md_federation}.


As a subclass of ONRA, KRORA problems can be modeled as instances of SDMU. MDP is a framework to formulate SDMU as a tuple $(\mathcal{S}, \mathcal{A}, \mathfrak{P}, \mathfrak{R})$, where $\mathcal{S}$ is the environment states set, $\mathcal{A}=\{\mathcal{A}(s)\ \forall s \in \mathcal{S}\}$, where $\mathcal{A}(s)$ is the set of the actions in state $s$. Let $\mathcal{S}'_{s,a}$ be the set of possible next states in the case of taking action $a$ in state $s$, $\mathfrak{P}(s,a,s')\!\!: \mathcal{S} \times \mathcal{A}(s) \times \mathcal{S}'_{s,a} \rightarrow [0,1]$ is the probability of the transition from $s$ to $s' \in \mathcal{S}'_{s,a}$, and $\mathfrak{R}(s,a)\!\!: \mathcal{S} \times \mathcal{A}(s) \rightarrow \mathbb{R}$ is the reward function. In KRORA, the service provider offers $\mathcal{I}=\{1,\ldots,I\}$ types of services, which are defined in the service catalog. Each state  $s$, as shown in Fig.~\ref{fig:krora}, corresponds to arrival of a service request wherein the RA takes an action. State is defined as $s=(\bm{c}, \bm{l}, \bm{d})$ where vector $\bm{c}$ represents the available network resources, $\bm{l}$ is the vector representing the active requests of each service type $i \in \mathcal{I}$, and $\bm{d}$ is a 0/1 vector identifying the service type of the current request. $\mathcal{A}(s)$ is the set of feasible different resource allocation patterns, e.g., different feasible paths for the request. $\mathfrak{R}(s,a)$ is the reward of each RA pattern, which is already known or can be determined. However, $\mathfrak{P}(s,a,s')$, as it is shown in Fig. \ref{fig:krora}, is determined by the arrival and departure rates of the request during period $t_{2}$, which are \textbf{not} known in advance. An example of such formulation is presented in Section \ref{sec:md_federation}.

RL is a common approach to solve MDP. In each state $s$, the agent takes action $a$ and, consequently, the environment determines the next state $s'$ and reward $\mathfrak{R}(s,a)$. The fundamental concept in RL is Temporal Difference (TD), i.e., the difference between the current value of the state-action pair $Q[s,a]$ and the new estimation obtained according to this interaction. This difference is used in a TD-update rule to improve the policy, e.g., in the Q-Learning, the TD-update rule is
\begin{equation*}
Q[s,a] = (1 - \alpha)Q[s,a] + \alpha \Big(\mathfrak{R}(s,a) + \gamma \max_{a'} Q[s',a']\Big),
\end{equation*}
where $\alpha$ is the learning rate and $\gamma$ is the discount factor \cite{sutton2018reinforcement}.



Despite successful applications of LR in contexts like games and robotics, there are concerns about the practicality of applying RL to ONRA due to the sample complexity issue. The RL agent can be trained either \emph{offline} or \emph{online}. In the offline approach, learning and evaluating are separated stages, where in the first stage, the agent learns the policy by exploring state space via a large number of trial-and-error interactions with the (simulated) environment but it is not evaluated; in the second stage, the learned policy is evaluated without learning. In the online approach, the agent should learn the policy in the real environment, so, it is evaluated while it is learning. 

The offline approach is applicable only if there is an isolated training environment that is not practical in KRORA due to the lack of information about future requests. Moreover, in the offline approach, the agent should be re-trained if the environment changes, which is quite often in KRORA due to changes in request rates\footnote{Please note that the approach composed of these steps: (1) learning the stochastic behavior of requests, (2) developing a simulator using the stochastic models, (3) training the agent offline in the simulated environment, (4) using the trained agent/policy in the real environment; is not a practical solution either, as it again needs (considerable) data for learning the stochastic behavior and also re-training the agent in non-stationary environments.}. So, the online training approach is more suitable in KRORA but due to the sample complexity issue, the agent's decisions are sub-optimal for a high number of requests that decrease the cumulative reward significantly. When request rates are non-stationary, the agent should \emph{continually learn} the policy, which makes the problem worse because, if dynamics of the environment change faster than the time it takes to find the optimal policy, then the policy is always sub-optimal.

In summary, online agent training is the practical approach to deal with KRORA via RL, but a solution is needed to alleviate the prolonged learning period. In the following sections, we propose the RADAR (\textbf{R}esource \textbf{A}llocation via mo\textbf{D}el le\textbf{AR}ning) framework built on model-based RL to improve sample efficiency of RL.

\vspace{-0.7mm}
\section{MBRL Framework for KRORA}
\label{sec:framework}
\vspace{-1.25mm}

\subsection{Design Concepts}
\vspace{-0.5mm}
In traditional model-free reinforcement learning, like Q-Learning and DQN, it is assumed that no information about the dynamics of the environment is known in advance. Thus, the agent has to learn the policy by a TD-update rule in a high number of real interactions with the environment, which leads to the sample complexity problem. While this is a reasonable assumption in some problems, we can do much better in KRORA, as some information about the environment is known in advance, and some information can also be learned via the real interactions. This is the design idea of the RADAR framework built on MBRL---in each real interaction with the environment, the agent updates the policy by a given TD-update rule exactly in the same way that model-free RL does; but moreover, it also learns a \emph{model} of the environment, and then exploits it to update the policy furthermore using samples generated by the model instead of relying only on the real interactions with the environment. So, it can  efficiently deal with the sample complexity issue by generating (guided) synthetic samples. However, to utilize this approach, a few fundamental questions should be answered: ($i$) what is the model of the environment? ($ii$) how to learn the model? ($iii$) how to integrate the model into the RL architecture? In the following, we answer these questions in RADAR.

In general, developing an analytical model for a (stochastic) environment that, for given $s$ and $a$, determines $\mathfrak{P}(s,a,s')$ $\forall s' \in \mathcal{S}'_{s,a}$ is not an easy task. Therefore, in RADAR, we resort to a \emph{sample} model that in each invocation, for given $s$ and $a$, returns just a $s' \in \mathcal{S}'_{s,a}$ and $\mathfrak{R}(s,a)$. Referring back to Fig. \ref{fig:krora}, the steps of this model are as follows: ($i$) action $a$ is applied in state $s=(\bm{c}, \bm{l}, \bm{d})$ that changes $\bm{c}$ and $\bm{l}$ to $\tilde{\bm{c}}$ and $\tilde{\bm{l}}$, ($ii$) the reward $\mathfrak{R}(s,a)$ is determined, ($iii$) sequentially, samples are taken from the arrival and departure stochastic processes. If the sample is departure, it is applied in the network, it changes $\tilde{\bm{c}}$ and $\tilde{\bm{l}}$ to $\bm{c}'$ and $\bm{l}'$, and another sample is taken; however, if the sample is arrival, it determines the next state $s'=(\bm{c}', \bm{l}', \bm{d}')$.

In KRORA, as the service provider knows the available network resources $\bm{c}$ and alive services $\bm{l}$, so can it determine the effect of each RA action for the given request for service type $i$; i.e., it can perform step ($i$). Moreover in KRORA, it is assumed that the immediate reward is known; so, step ($ii$) is straightforward. The only remaining part is the requests arrival/departure \emph{stochastic processes}, which are not known in advance. In the RADAR framework, these processes are \emph{learned} from the interactions with the real environment. There are various approaches to learn a stochastic process from data \cite{koller1998using,foong2020meta}. In RADAR, it is a black-box which on one hand, takes information about the arrival/departure times of real requests, and internally learns a stochastic process for it; on the other hand, it can generate samples of the stochastic process. So, until now, we answered the first two questions---the model is a sample model and model learning is indeed learning the requests arrival/departure stochastic processes. 


The last issue is the integration of the sample model in the RL architecture to improve sample efficiency by updating the policy using the synthetic interactions by the model. In RADAR, this can be conducted in both time intervals $t_{1}$ and/or $t_{2}$ depicted in Fig. \ref{fig:krora}.
In $t_{1}$, the request is known, so we seek the best action via \emph{decision-time planning} composed of exploration and exploitation steps, but in $t_{2}$ the action has already been taken, so we aim to improve the policy for the next unknown requests by \emph{background exploration} of the MDP.

To explore state space, either in background or decision-time, we start from a given state and update the policy by the RL TD-update rule in a number of trajectories of synthetic samples where, to emphasize exploration, random actions are taken. In exploitation, we update the value of each action in the given state by generating a number of synthetic samples, but to exploit the policy, we greedily select the action with the highest value. Details are elaborated in the following.


\vspace{-0.5mm}
\subsection{RADAR Framework Architecture}
\begin{figure}[t]
\begin{center}
\centerline{\includegraphics[width=0.97\linewidth]{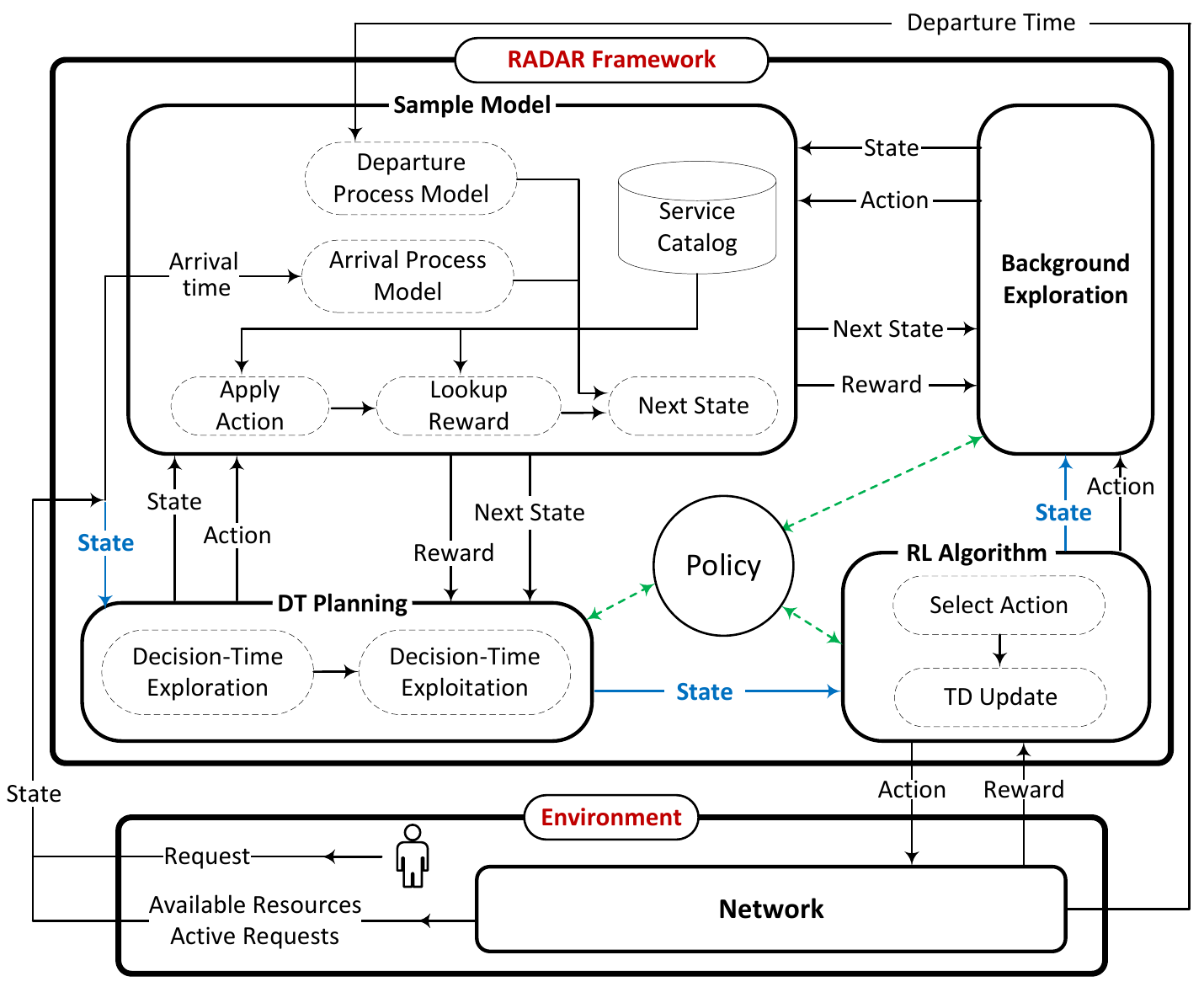}}
\vspace{-3mm}
\caption{The architecture of the RADAR framework}
\label{fig:framework}
\vspace{-6.5mm}
\end{center}
\end{figure}

The architecture of RADAR built on the explained solutions, is shown in Fig. \ref{fig:framework}. The user's request in combination with the available resources of the network and alive services are given as a new state to the framework. The arrival time is used to update the model of the arrival stochastic process. The state itself is processed by the decision-time planning module. The components of the module utilize the sample model and update the policy. Then, the state is given to the RL algorithm, which selects the best action according to its exploration strategy. The action is given to the environment and, after a while, when the new state is given, the policy will be updated by the RL TD-update rule. During this time, i.e., $t_{2}$ in Fig. \ref{fig:krora}, the state and the selected action are processed by the background exploration module, which also uses the sample model to update the policy. Finally, when a service departs the network, its departure time is used to update the departure stochastic model. The sample model module, besides the arrival and departure models, uses the service catalog to determine the $\mathfrak{R}(s, a)$ and $s'$ for each given $s$ and $a$.

\vspace{-0.5mm}

RADAR is quite flexible;  according to the constraints on $t_{1}$ and $t_{2}$, not only the number and length of the exploration trajectories can be adjusted per module but also each module can be enabled/disabled independently. Table \ref{table:MBRLs} shows four RL algorithms (out of 8 possible options) as representative solutions based on RADAR to be used in different practical situations. \textsf{MFRL} is the standard model-free RL where all modules are disabled. \textsf{MB-BGEX} is the algorithm that can be used in real-time KRORA problems where $t_{1} = 0$. If the state instantly changes after taking action, i.e., $t_{2} = 0$, the \textsf{MB-DTP} can be used. Finally, \textsf{MB-Full} can be used when $t_{1}, t_{2} > 0$.

\begin{table}[t] 
\begin{center}
\caption{Algorithms based on the RADAR Framework}
\vspace{-5mm}
\label{table:MBRLs}
\centering
\small\addtolength{\tabcolsep}{0pt}
\scalebox{0.87}{
\begin{tabular}{|c|c|c|c|l|}
  \hline
  Algorithm & BG      & \multicolumn{2}{c|}{Decision-Time}  & \hspace{1cm} Comments \\
  \cline{3-4}
            & Explore & Explore & Exploit &  \\
  \hline
  \hline
  \textsf{MFRL} & $\times$ & $\times$ & $\times$ & Standard model-free RL \\
  \hline
  \textsf{MB-BGEX} & $\checkmark$ & $\times$ & $\times$ & Real-time RA ($t_{1} = 0$)\\
  \hline
  \textsf{MB-DTP} & $\times$ & $\checkmark$ & $\checkmark$ & Instance next state ($t_{2} = 0$) \\
  \hline
  \textsf{MB-Full} & $\checkmark$ & $\checkmark$ & $\checkmark$ & $t_{1}, t_{2} > 0$ \\
  \hline
\end{tabular}	
}		
\end{center}
\vspace{-2mm}
\end{table}

\vspace{-1.5mm}
\subsection{RADAR Framework Implementation}
\vspace{-0.5mm}

The main function of the sample model module is the step procedure, depicted in Algorithm \ref{alg:sample_model_step}, that returns $s'$ and $\mathfrak{R}(s,a)$ for a given $s$ and $a$. It first looks up the new request in the service catalog and finds the associated reward and required resources. Then, it applies the given action that updates $\bm{c}$ and $\bm{l}$. In lines 5 and 6, the next request is determined by sampling the arrival stochastic processes. Finally, according to the departure stochastic process samples, if an active service terminates before the next arrival, its resources are released.

Exploration, either in background time $t_{2}$ or decision-time $t_{1}$, is conducted by the \textsc{Explore} procedure shown in Algorithm \ref{alg:explore}. It explores the state space in $\theta$ trajectories starting from state $s_{0}$ with length $\kappa$. In the beginning, the sample model is initialized by the state $s_{0}$. Then, in each trajectory, if the action in state $s_{0}$ is already known, the policy is fixed in this state in line 3. As the objective is to explore the state space, a random action is selected in each state. The next state and the corresponding reward are given by the sample model; and finally, in line 8, the policy is updated. The background exploration in state $s$, where the action $a$ has already been taken, is implemented by invoking \textsc{Explore}$(s, a, \theta, \kappa)$; and the decision-time exploration in state $s$ is indeed \textsc{Explore}$(s$, NULL, $\theta, \kappa)$ as the action in this state is not known in $t_{1}$.

\begin{algorithm}[h]
\caption{\textsc{SampleModelStep}$(s=(\bm{c}, \bm{l}, \bm{d}),a)$}
\label{alg:sample_model_step}
\begin{spacing}{1.1}
\begin{small}
\begin{algorithmic}[1]
    \State $\mathfrak{R}(s,a) \gets $ lookup reward of new request $\bm{d}$ in service catalog
    \State $\bm{u} \gets $ lookup required resources of request $\bm{d}$ in service catalog
    \State $\tilde{\bm{c}} \gets $ apply action $a$ and resources $\bm{u}$ on $\bm{c}$
    \State $\tilde{\bm{l}} \gets $ apply action $a$ on $\bm{l}$
    \State $\mathcal{T} = \{t_{a,1},\ldots t_{a,I}\} \gets $ sample of arrival process $\forall i \in \mathcal{I}$
    \State $\bm{d}' \gets \argmin \mathcal{T}$, $t_{a} \gets \min \mathcal{T}$
    \For {$j \in \mathcal{I}$ s.t. $d_{j} > 0$}
       \State $t_{d,j} \gets $ sample of departure stochastic process $j$
        \State Release request $j$, update $\tilde{\bm{l}}$ and $\tilde{\bm{c}}$ \textbf{ if} {$t_{d,j} < t_{a}$}
    \EndFor
    \State $s' = (\tilde{\bm{c}}, \tilde{\bm{l}}, \bm{d}')$
	\State \Return $s'$, $\mathfrak{R}(s,a)$
\end{algorithmic}
\end{small}
\end{spacing}
\end{algorithm}
\begin{algorithm}[H]
\caption{\textsc{Explore}$(s_{0}, a_{0}, \theta, \kappa)$}
\label{alg:explore}
\begin{spacing}{1.1}
\begin{small}
\begin{algorithmic}[1]
    \For {$\theta$ times}
        \State Initialize the sample model by the given state $s_{0}$
        \State $\mathcal{A}(s_{0}) \gets \{a_{0}\}$ \textbf{if} {$a_{0} \neq $ NULL}  
        \State $s \gets s_{0}$
        \For {$\kappa$ times}
            \State $a \gets $ random action of $\mathcal{A}(s)$
            \State $s', \mathfrak{R}(s,a) \gets $ \textsc{SampleModelStep}$(s,a)$
            \State \textsc{TD-update}$(s,a,s',\mathfrak{R}(s,a))$
            \State $s' \gets s$
        \EndFor
    \EndFor
\end{algorithmic}
\end{small}
\end{spacing}
\end{algorithm}
\begin{algorithm}[H]
\caption{\textsc{Exploit}$(s_{0}, \theta, \kappa)$}
\label{alg:exploit}
\begin{spacing}{1.1}
\begin{small}
\begin{algorithmic}[1]
    \For {$\bar{a} \in \mathcal{A}(s_{0})$}
        \For {$\theta$ times}
            \State Initialize the sample model by the given state $s_{0}$
            \State $\mathcal{A}(s_{0}) \gets \{\bar{a}\}$, $s \gets s_{0}$
            \For {$\kappa$ times}
                \State $a \gets $ the highest value action in $\mathcal{A}(s)$
                \State $s', \mathfrak{R}(s,a) \gets $ \textsc{SampleModelStep}$(s,a)$
                \State \texttt{stack.push}$(s,a,s',\mathfrak{R}(s,a))$
                \State $s' \gets s$
            \EndFor
            \While{stack is not empty}
                \State $s,a,s',\mathfrak{R}(s,a) \gets $ \texttt{stack.pop()}
                \State \textsc{TD-update}$(s,a,s',\mathfrak{R}(s,a))$
            \EndWhile
        \EndFor
    \EndFor
\end{algorithmic}
\end{small}
\end{spacing}
\end{algorithm}

In the decision-time exploitation, we aim to evaluate the quality of each action, so we generate a number of trajectories per action in the state to update the value of the action according to the value of the next states. Consider a trajectory $(s_{0},a_{0})$ $\rightarrow$ $(s_{1},a_{1})$ $\rightarrow$ $\ldots$ $\rightarrow$ $(s_{\kappa},a_{\kappa})$. In the standard operation of RL algorithms, value $(s_{0}, a_{0})$ is updated according to value $s_{1}$ and \emph{then} the value of $(s_{1},a_{1})$ is updated. So, action value updates in states $s_{1}$ \ldots $s_{\kappa}$ do not have any effect on the value $(s_{0}, a_{0})$ in this trajectory. To take these updates into account, in the decision-time exploration, we update the values \emph{backward}, i.e., from $(s_{\kappa}, a_{\kappa})$ to $(s_{0}, a_{0})$. The details of the implementation is shown in Algorithm \ref{alg:exploit}. To exploit the current policy, the action with the highest value is selected in line 6; and the backward updates are implemented using a stack.

\vspace{-2.0mm}
\section{Multi-Domain Service Federation}
\label{sec:md_federation}
\vspace{-1.5mm}
In this section, as a proof of concept, we apply RADAR on MDSF \cite{bakhshi2021globe} that enables the service provider to collaborate with other providers in service provisioning where the federation contract
provides extra resources for the consumer domain at the cost of federation. In MDSF, the RA either determines the domain to deploy the given request or rejects it while it is not aware of future requests; i.e., $\mathcal{A}(s)$=\{\textsc{reject}, \textsc{local deploy}, \textsc{federate}\}. Here, we consider MDSF between a consumer domain and a provider domain. Each service type $i \in \mathcal{I}$ is specified by a tuple $(w_{i},r_{i})$ in the service catalog, where $w_{i}$ is the total amount of required resources e.g., CPU cores, and $r_{i}$ is the revenue if a request of type $i$ is accepted; moreover, according to the federation contract, $\varphi_{i}$ is the cost of deploying request $\delta_{i}$ of type $i$ in the provider domain. So, $\mathfrak{R}(s,a)$ of the \textsc{reject}, \textsc{local deploy}, and \textsc{federate} actions are respectively known as 0, $r_{i}$, $r_{i} - \varphi_{i}$. The objective is to maximizes the service provider's average profit 
\begin{equation}
\label{eq:profit}
\frac{1}{|\mathcal{D}|}\sum_{i \in \mathcal{I}}\Big(\sum_{\delta_{i} \in \mathcal{L}}  r_{i} + \sum_{\delta_{i} \in \mathcal{F}} (r_{i} - \varphi_{i})\Big);
\end{equation}   
where, $\mathcal{D}$ is the set of requests; and $\mathcal{L}$ ($\mathcal{F}$) is the set of the services deployed in the consumer (provider) domain. It is easy to show that MDSF is a KRORA problem.

In MDSF, the state is defined as $s=(\bm{c}_{c}, \bm{c}_{p}, \bm{l}_{c}, \bm{l}_{p}, \bm{d})$, where $\bm{c}_{c}$ and $\bm{c}_{p}$ are respectively the current available capacity of the consumer and provider domains. $\bm{l}_{c}$ and $\bm{l}_{p}$ are vectors wherein the $i$-th element is the number of alive services of type $i$ in the consumer domain and in the provider domain, respectively. 

To solve MDSF through RADAR, we need to specify the RL algorithm and the arrival/departure stochastic process models; the remaining components of the framework are problem-independent. In \cite{bakhshi2021globe}, we showed that average-reward reinforcement learning, i.e., the R-Learning algorithm \cite{schwartz1993reinforcement}, is an efficient solution for the problem. In this algorithm, in addition to the values of the actions $Q[s,a]$, a parameter $\rho$, which is the average reward of the MDP, is also learned. So the TD-update rule of the RL algorithm is:
\begin{equation*}
Q[s,a] \gets (1 - \alpha)Q[s,a] + \alpha \Big(\mathfrak{R}(s,a) - \rho +  \max_{a'} Q[s',a'] \Big),
\end{equation*}
and
\begin{equation*}
\rho \gets (1 - \beta) \rho + \beta \Big(\mathfrak{R}(s,a) - \max_{a}Q[s,a] + \max_{a'}Q[s',a'] \Big).
\end{equation*}

For the arrival and departure stochastic processes, we assumed that they are Poisson processes, as usually done in the literature. Under this assumption, we only need to learn the expected value of the distribution. Here, we use exponential moving average to estimate it. It must be noted that any other method can also be used to learn stochastic processes. In section \ref{sec:inaccuracy}, we evaluate the effect of this assumption.

\vspace{-0.25mm}
\section{Simulation Results}
\label{sec:results}
\vspace{-0.75mm}
In this section, we evaluate the RADAR-based RA algorithms, shown in Table \ref{table:MBRLs}, in the MDSF problem. As the \emph{theoretical} performance bound, the results of the offline R-Learning algorithm, shown is near-optimal \cite{bakhshi2021globe}, are also included by legend \textsf{RL-Offline}. In \textsf{RL-Offline}, the agent is trained offline using $\theta$ = $10^3$ trajectories with $\kappa$ = $10^4$ requests per trajectory. The performance metric is the average profit defined in \eqref{eq:profit}, and the capacity of the consumer and provider domains are respectively 100 and 50 CPU cores. The default settings of the simulated service types are shown in Table \ref{table:services}. The following results are the average of 20 experiments.

\vspace{-0.75mm}
\subsection{Exploration Breadth vs. Depth}
\begin{figure}[t]
\begin{center}
\vspace{-8mm}
    \subfloat[Background Exploration]{
        \includegraphics[width=0.49\linewidth]{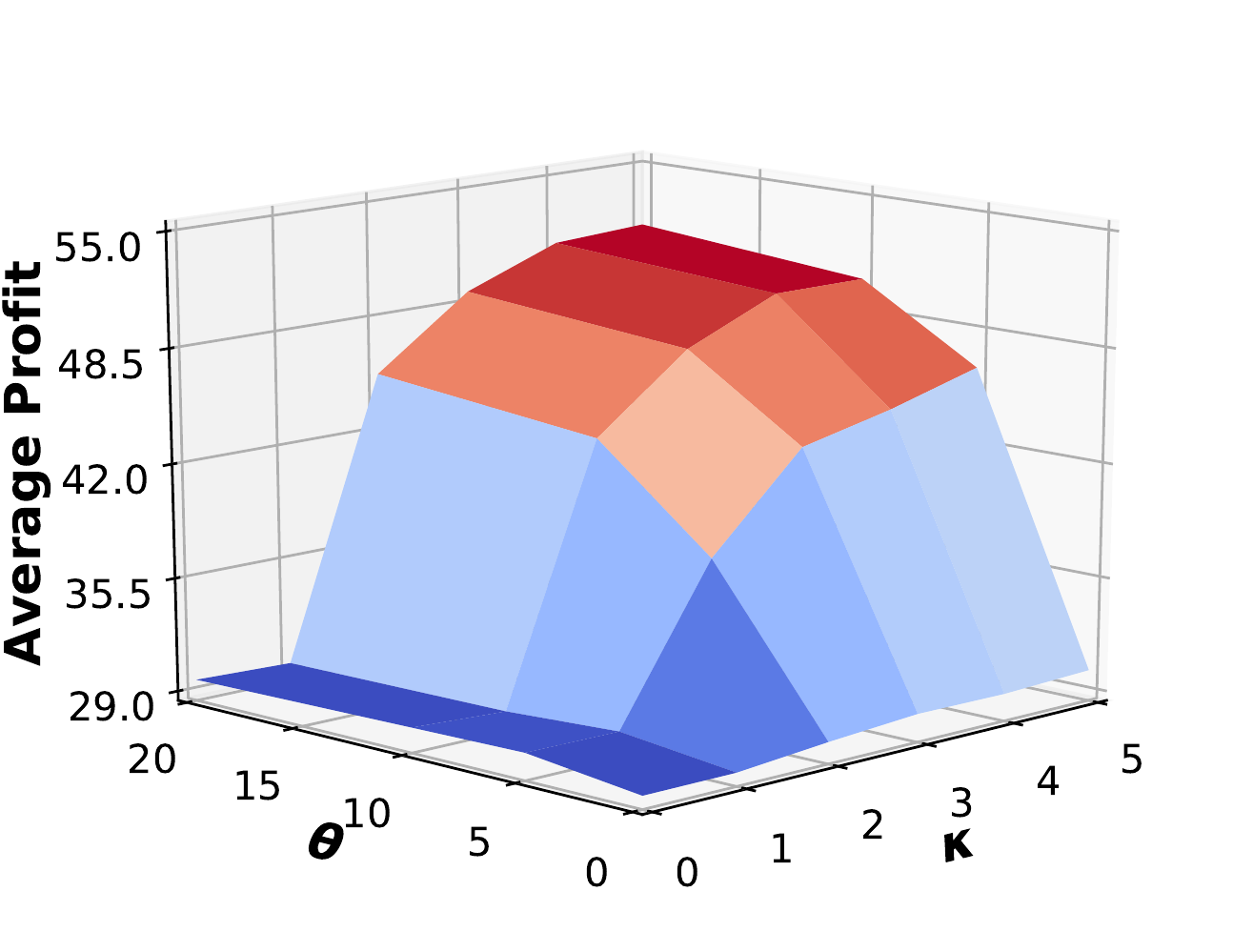}
    }
    \subfloat[Decision-Time Exploitation]{
        \includegraphics[width=0.49\linewidth]{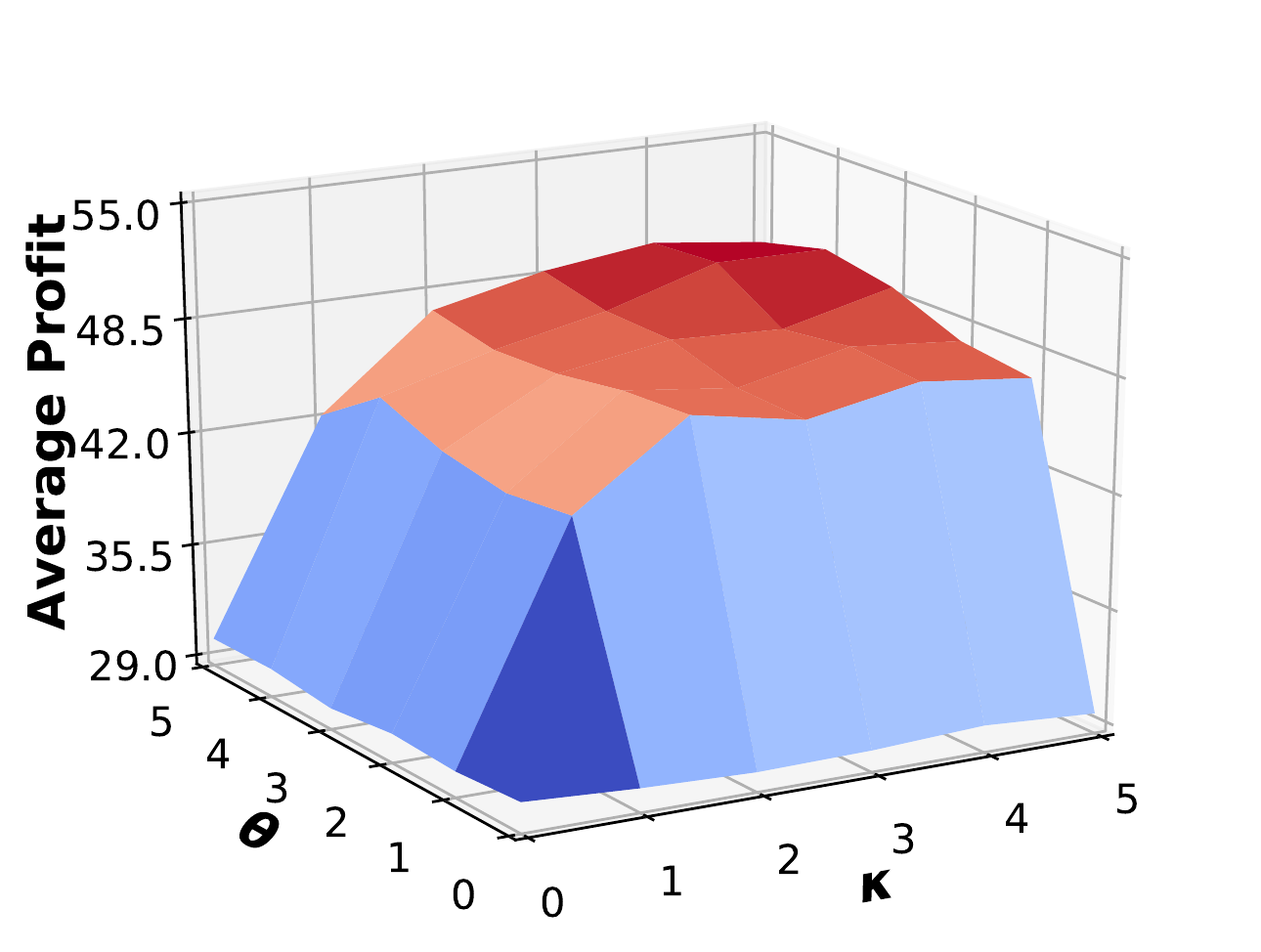}
    }
\vspace{-1mm}
\caption{The effect of $\theta$ and $\kappa$ on the performance of each modules}
\label{fig:theta_kappa}
\vspace{-1.5mm}
\end{center}
\end{figure}

\vspace{-0.99mm}
In the RADAR's modules, the parameters $\theta$ and $\kappa$ determine the breadth and depth of exploration trajectories. The effects of these parameters on the framework modules' performance are shown in Fig. \ref{fig:theta_kappa} (the result of the decision-time exploration is similar; omitted due to space limitations). While evaluating each module, the other modules are disabled. It is seen that $\theta$ and $\kappa$ should be larger than a threshold to effectively explore the state space. The background exploration needs larger $\theta$  as due to the unknown next state, it needs more trajectories to visit possible next states.  Increasing the values of the parameters, which increases the solution complexity, beyond certain values does not yield better performance. Based on these results, in the following simulations, we set $\theta$ = 5 and $\kappa$ = 3 in background exploration, $\theta$ = 3 and $\kappa$ = 2 in decision-time exploration, and $\theta$ = 1, $\kappa$ = 3 in decision-time exploitation.  

\vspace{-0.75mm}
\subsection{Learning Capability}
\begin{figure*}
\centering
\begin{minipage}{.32\linewidth}
  \includegraphics[width=\linewidth]{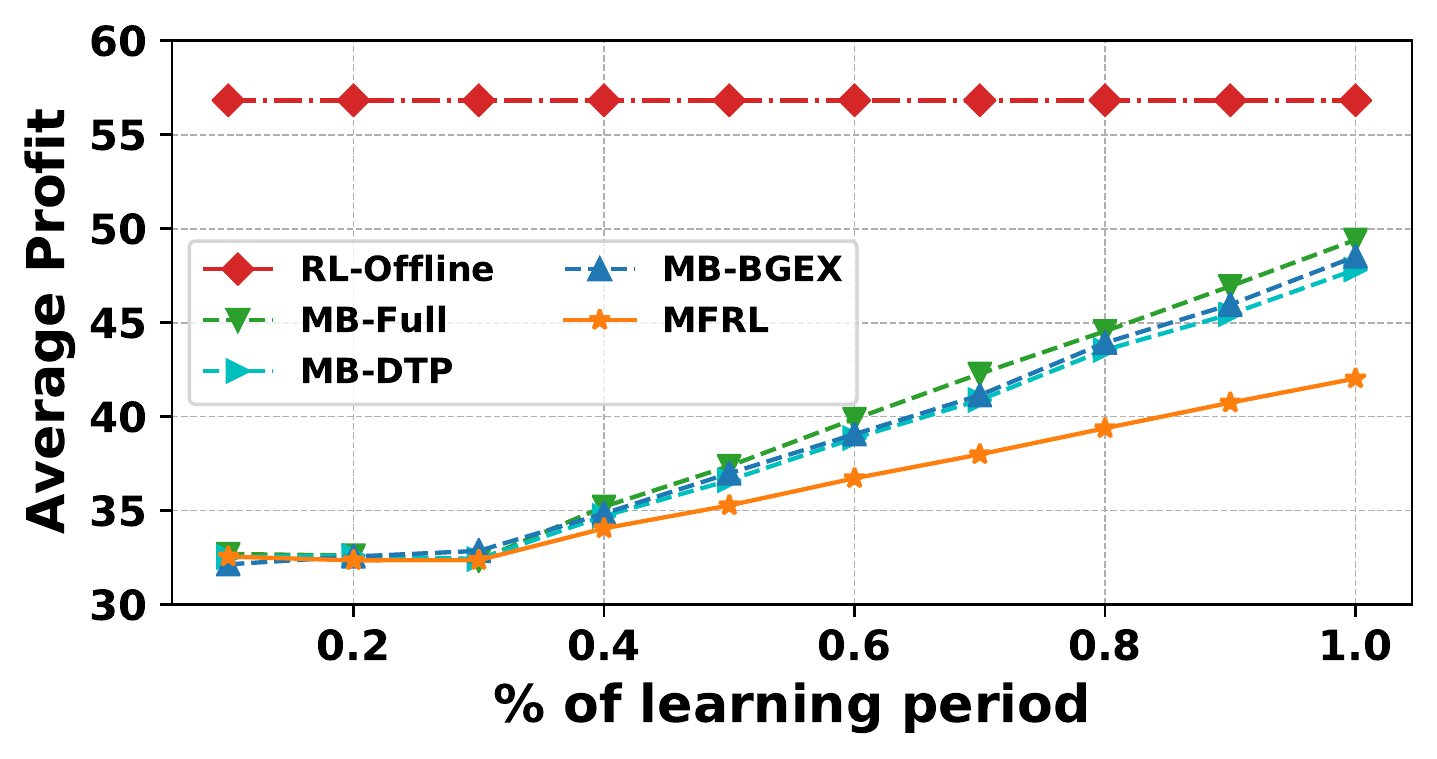}
  \vspace{-8mm}
  \caption{Effect of length of learning interval}
  \label{fig:learn_interval}
\end{minipage}
\vspace{-1.5mm}
\hfill
\begin{minipage}{.32\linewidth}
  \includegraphics[width=\linewidth]{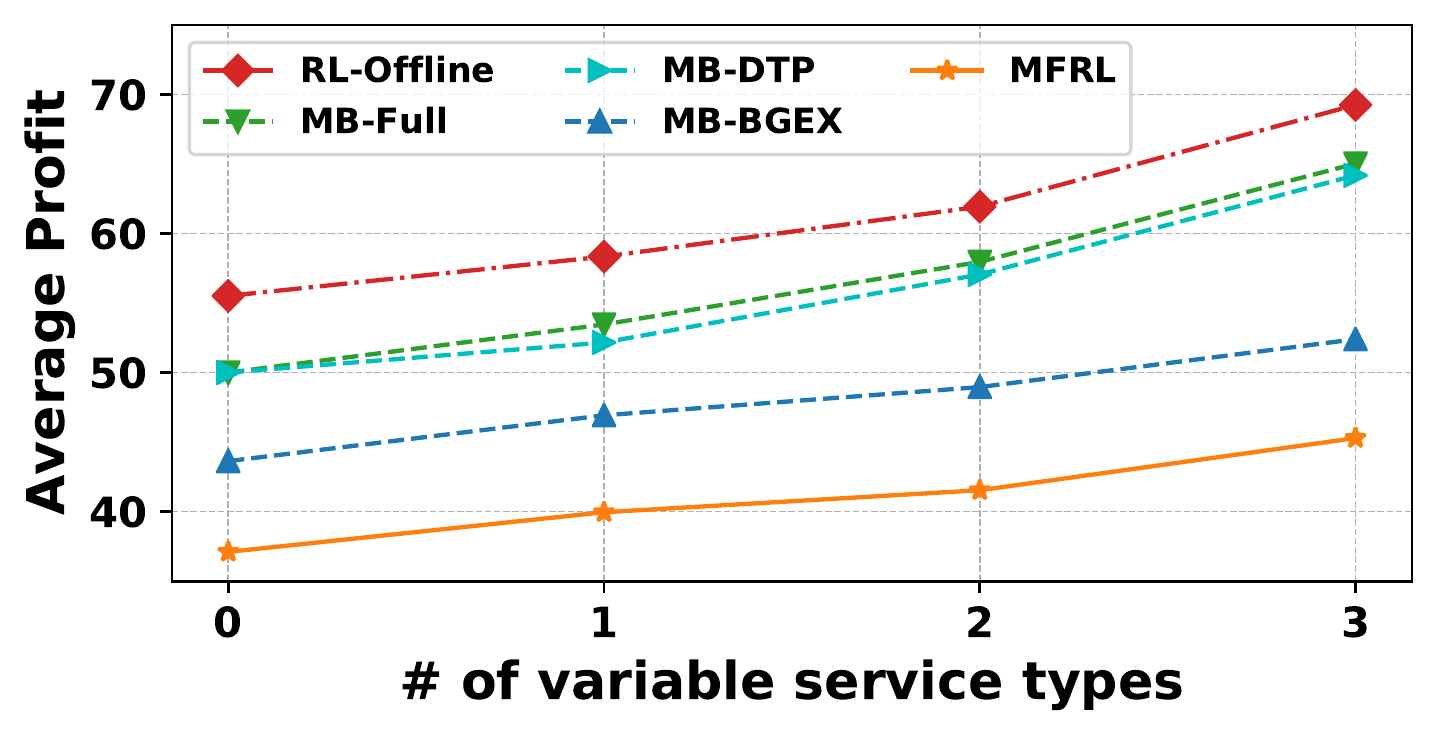}
  \vspace{-8mm}
  \caption{Effect of variable traffic}
  \label{fig:var_traffic}
\end{minipage}
\vspace{-1.5mm}
\hfill
\begin{minipage}{.32\linewidth}
  \includegraphics[width=\linewidth]{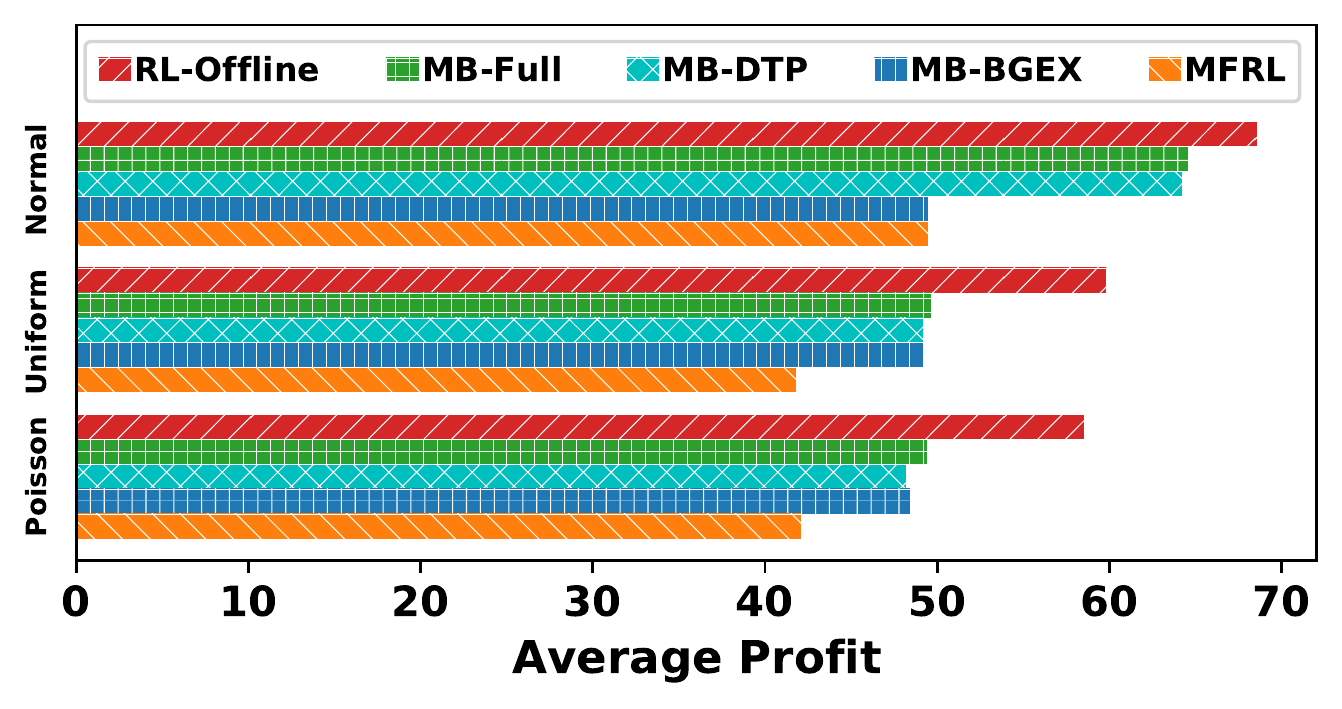}
  \vspace{-8mm}
  \caption{Effect of inaccurate traffic model}
  \label{fig:traffic_dist}
\end{minipage}
\vspace{-1.5mm}
\end{figure*}

\vspace{-0.99mm}
In this section, we evaluate how fast and efficient the RADAR-based algorithms find the policy utilizing the sample model. For this assessment, learning of the sample model and also policy update are disabled after a given time period from simulation start time. Fig. \ref{fig:learn_interval} shows the performance of the algorithms with respect to the length of the learning period (in terms of percentage of simulation time). For the learning period less than 30\% of the simulation time, all algorithms are almost the same, and model-based RL does not perform better than model-free RL because, in this case, the algorithms do not have sufficient opportunity to explore the huge state space of the MDP, i.e., $|\mathcal{S}| = O(10^9)$. However, by providing more opportunities for learning, the MBRL algorithms learn a better policy faster, which leads to a higher average profit than \textsf{MFRL}. These results also show the sample efficiency of RADAR; \textsf{MBRL-Full}, which is learning during the whole simulation time using about 16900 real samples, achieves an 11\% optimality gap compared to \textsf{RL-Offline}, which was trained with $10^7$ samples; i.e., 590 times fewer samples.

\begin{table}[t] 
\begin{center}
\vspace{0mm}
\caption{Simulated Service Type Parameters}
\vspace{-2mm}
\label{table:services}
\centering
\small\addtolength{\tabcolsep}{0pt}
\scalebox{0.9}{
\begin{tabular}{|c|c|c|c|c|c|}
  \hline
  $i$ & $w_{i}$ (core) & $r_{i}$ (\$) & $\varphi_{i}$ (\$) & $\lambda_{i}$ (req\,$/$\,h)& $\mu_{i}$ (req\,$/$\,h)\\
  \hline
  \hline
  1 & 2 & 100 & 30 & 10 & 0.4 \\
  \hline
  2 & 1 & 20 & 5 & 5 & 0.05 \\
  \hline
  3 & 3 & 50 & 45 & 2 & 0.2 \\
  \hline
\end{tabular}	
}		
\end{center}
\vspace{-2mm}
\end{table}

\vspace{-0.75mm}
\subsection{Adaptiveness Capability}
\vspace{-0.99mm}
As discussed, one of the advantages of MBRL is its continual learning capability to adapt the policy to the changes in the environment. In this section, we evaluate the performance of the RADAR-based algorithms in the case that arrival rates of requests are changing over time. For this purpose, the simulation time is divided into 5 equal time slots, and if a service type is variable, then $\lambda_{1}= [6, 8, 10, 8, 6]$, $\lambda_{2} = [3, 1,  3, 1, 3]$, and $\lambda_{3}= [4, 2, 0, 2, 4]$ req/h during the time slots. Fig. \ref{fig:var_traffic} shows the average profit of the algorithms w.r.t. the number of variable service types. As it is seen, \textsf{MB-Full} considerably outperforms \textsf{MFRL} that means RADAR can improve the performance up to 44\% in comparison to the standard model-free RL. Moreover, the gap between \textsf{MB-Full}, using about $4 \times 10^5$ real samples, and \textsf{RL-Offline}, trained offline with  $10^7$ samples, is independent of the number of the variable service types, hence showing the excellent sample efficiency of RADAR and its capability to continually learn the sample model and use it to adapt the policy with the changes of the environment.

\vspace{-0.75mm}
\subsection{Model Inaccuracy Tolerance}
\label{sec:inaccuracy}
\vspace{-0.99mm}
As mentioned, to apply RADAR on MDSF, we assumed that the arrival and departure of requests are Poisson processes. In this section, we evaluate the capability of the RADAR-based solutions in tolerating the inaccuracy of the model. The average profits of the algorithms w.r.t. different stochastic processes are shown in Fig. \ref{fig:traffic_dist}. In the Poisson process, the Inter-Arrival Time (IAT) and Holding Time (HT) are respectively exponential distributions with mean $1 / \lambda_{i}$ and $1 / \mu_{i}$, shown in Table \ref{table:services}. In the case of uniform distribution, IAT and HT are uniformly distributed in intervals $[0, 2 / \lambda_{i}]$ and $[0, 2 / \mu_{i}]$, and for the normal distribution, IAT = $\max(0, \mathcal{N}(10/\lambda_{i}, (5/\lambda_{i})^2)$ and HT = $\max(0, \mathcal{N}(10/\lambda_{i}, (5/\lambda_{i})^2)$.  These results show that modeling the normal and uniform distributions as a Poisson process does not significantly affect the performance; i.e., the sample model learned by RADAR and the corresponding synthetic samples can remarkably improve the performance. 

\vspace{-1mm}
\section{Conclusion}
\label{sec:conclusion}
\vspace{-1.25mm}
In this paper, for KRORA problems, where the immediate rewards of  RA's actions are known, we developed the RADAR framework built on MBRL, which is capable of continual learning. In combination with the information about the network resources, which the service provider has, the known rewards are utilized to develop a sample model of the environment. In this model, the arrival and departure stochastic processes of service requests are learned via interactions with the environment. The model is integrated with an RL algorithm to generate synthetic samples used in both background and decision-time for updating the policy. Evaluation of RADAR in the MDSF problem showed that it is not only applicable to KRORA problems, but also it can improve the average profit up to 44\% compared to the standard model-free RL.

\bibliographystyle{ieeetr}
\bibliography{refs}

\end{document}